\documentclass[prb,twocolumn,floatfix,superscriptaddress,amsmath,citeautoscript,aps,longbibliography]{revtex4-2}
\pdfoutput=1
\usepackage[T1]{fontenc}\usepackage[latin1]{inputenc}
\usepackage{dcolumn,bm,graphicx,xcolor,booktabs,microtype,afterpage} \graphicspath{{figs/}}
\usepackage[charter,greekuppercase=italicized]{mathdesign}
\renewcommand{\figurename}{Fig.}
\renewcommand{\tablename}{Table}
\makeatletter\renewcommand{\fnum@figure}[1]{\figurename~\thefigure.}\makeatother
\makeatletter\renewcommand{\fnum@table}[1]{\tablename~\thetable.}\makeatother
\newcount\hh \newcount\mm
\hh=\time \divide\hh by 60
\mm=\hh \multiply\mm by 60 \mm=-\mm
\advance\mm by \time
\def\now{\number\hh:\ifnum\mm<10{}0\fi\number\mm}
\usepackage[colorlinks,plainpages=false,linkcolor=blue,urlcolor=blue,citecolor=blue,pdfpagemode=UseNone,pdfstartview=FitBH]{hyperref}
\usepackage[version=4]{mhchem}

\begin{document}

\title{Phonon softening and atomic modulations in EuAl$_4$}

%\author{A.~N.~Korshunov}
%\affiliation{European Synchrotron Radiation Facility (ESRF), BP 220, F-38043 Grenoble Cedex, France}
%\author{A.~S.~Sukhanov}
%\affiliation{Institut f{\"u}r Festk{\"o}rper- und Materialphysik, Technische Universit{\"a}t Dresden, D-01069 Dresden, Germany}
\author{A.~N.~Korshunov, A.~S.~Sukhanov}
\affiliation{Institut f{\"u}r Festk{\"o}rper- und Materialphysik, Technische Universit{\"a}t Dresden, D-01069 Dresden, Germany}
\author{S.~Gebel}
\affiliation{Institut f{\"u}r Festk{\"o}rper- und Materialphysik, Technische Universit{\"a}t Dresden, D-01069 Dresden, Germany}
\affiliation{Brazilian Synchrotron Light Laboratory (LNLS), Brazilian Center for Research in Energy
and Materials (CNPEM), Campinas, 13083-970 Sao Paulo, Brazil}
\author{M.~S.~Pavlovskii}
\affiliation{Kirensky Institute of Physics, Siberian Branch, Russian Academy of Sciences, Krasnoyarsk 660036, Russian Federation}
\author{N.~D.~Andriushin}
\affiliation{Institut f{\"u}r Festk{\"o}rper- und Materialphysik, Technische Universit{\"a}t Dresden, D-01069 Dresden, Germany}
\author{Y.~Gao}
\affiliation{Department of Physics and Astronomy, Rice University, Houston, Texas 77005, USA}
\author{J.~M.~Moya}
\thanks{Current affiliation: Department of Chemistry, Princeton University, Princeton, New Jersey 08544, USA}
\affiliation{Department of Physics and Astronomy, Rice University, Houston, Texas 77005, USA}
\affiliation{Rice Center for Quantum Materials (RCQM), Rice University, Houston, Texas 77005, USA}
\affiliation{Applied Physics Graduate Program, Smalley-Curl Institute, Rice University, Houston, Texas 77005, USA}
\author{E.~Morosan}
\affiliation{Department of Physics and Astronomy, Rice University, Houston, Texas 77005, USA}
\author{M.~C.~Rahn}
\thanks{Corresponding author: marein.rahn@tu-dresden.de}
\affiliation{Institut f{\"u}r Festk{\"o}rper- und Materialphysik, Technische Universit{\"a}t Dresden, D-01069 Dresden, Germany}

\begin{abstract} 
$\ce{EuAl4}$ is a rare earth intermetallic in which competing itinerant and/or indirect exchange mechanisms give rise to a complex magnetic phase diagram, including a centrosymmetric skyrmion lattice. These phenomena arise not in the tetragonal parent structure but in the presence of a charge density wave (CDW), which lowers the crystal symmetry and renormalizes the electronic structure. Microscopic knowledge of the corresponding atomic modulations and their driving mechanism is a prerequisite for a deeper understanding of the resulting equilibrium of electronic correlations and how it might be manipulated. Here, we use synchrotron single-crystal X-ray diffraction, inelastic X-ray scattering, and lattice dynamics calculations to clarify the origin of the CDW in $\ce{EuAl4}$. We observe a broad softening of a transverse acoustic phonon mode that sets in well above room temperature and, at $T_{\text{CDW}}=142$~K, freezes out in an atomic displacement mode described by the superspace group $Immm(00\gamma)s00$. In the context of previous work, our observation is a clear confirmation that the CDW in $\ce{EuAl4}$ is driven by electron-phonon coupling. This result is relevant for a wider family of BaAl$_4$ and ThCr$_2$Si$_2$-type rare-earth intermetallics known to combine CDW instabilities and complex magnetism.
\end{abstract}

\maketitle

\section{Introduction}

Electronic instabilities in the form of charge-density waves (CDWs) have recently gained increased attention. This is due to their prevalence in various topical platforms of quantum matter. CDW order has been found to coexist and/or couple to diverse emergent phases, whether in transition metal dichalcogenides~\cite{Ritschel2015}, cuprate and iron-based unconventional superconductors \cite{Loret:2019aa,Silva-Neto:2014aa}, or in systems with complex (topological) forms of magnetic order~\cite{Yasui:2020aa,Teng:2023aa,Chen:2016aa}. Crucially, the ability to induce, suppress, or modify CDWs could be a powerful means to tune emergent order parameters of correlated electron systems. Bottom-up insights into the microscopic mechanisms that underlie the formation of CDWs in pertinent materials are, therefore, of great interest.

A familiar CDW scenario is the Peierls instability of a one-dimensional metallic chain with one electron per site. Here, the perfect Fermi surface nesting (FSN) for Umklapp scattering with momentum transfers $\mathbf{Q}=2\,\mathbf{k}_\mathrm{F}$ favours the formation of a commensurate superstructure. Backfolding of the bands opens up a CDW gap, which implies a metal-insulator transition (MIT). Coupling to the lattice and the divergence in the real part of the Lindhard electronic response function translate to a Kohn anomaly (softening) in the phonon spectrum and, eventually, the formation of a periodic structural distortion at $\mathbf{q}_\mathrm{CDW}=2\,\mathbf{k}_\mathrm{F}$.

However, even among real one-dimensional materials, the idealized conditions of Peierls picture are rarely realized. CDW phases that have been studied in topical two- and three-dimensional systems bear even less resemblance to the limiting case of perfect FSN~\cite{Johannes:2008aa,Zhu:2017aa}. Instead, momentum-dependent electron-phonon coupling (EPC) may take on the dominant role. For instance, in the layered quasi-2D CDW material \ce{NbSe2}, no FSN is present \cite{Xi:2015aa}. Consequently, the system avoids an MIT -- while the phonon spectrum still develops a Kohn anomaly and the associated softening at $\mathbf{q}_\mathrm{CDW}$. It has been shown that this form of CDW originates in the momentum dependence of the EPC matrix element that couples occupied and unoccupied electronic states with the momentum and energy of the critical lattice fluctuation~\cite{Zhu:2015aa}.

Finally, three-dimensional materials have been known to feature more complex CDW mechanisms involving electronic correlations (i.e. spin and orbital degrees of freedom and their exchange interactions). These systems may contradict the characteristics of conventional CDWs, e.g. they may show FSN in the absence of both phonon softening and MIT~\cite{Zhu:2015aa}. In clarifying such scenarios, momentum- and energy-resolved evidence from complementary techniques that probe the chemical structure, lattice fluctuations, and the electronic structure becomes pivotal \cite{Korshunov:2023aa,Zhu:2017aa}.

The CDW system \ce{EuAl4} has recently gained special attention because of its potential for non-trivial topology, both in real- and momentum space~\cite{Ramakrishnan:gq5015,Shang:2021aa,Shimomura:2018aa,Kaneko:2021aa,Moya:2022aa}. The electronic structure of this material features a Dirac point (albeit $\approx200$\,meV above the Fermi surface) and is closely related to the Weyl nodal ring semimetal EuGa$_4$~\cite{Lei:2023aa}. At $T_\mathrm{N}=15.4$\,K, the large localized magnetic moments (\ce{Eu^2+}, $S=7/2$, $L=0$) order and open up a complex magnetic phase diagram, including three re-ordering transitions at zero field and a sequence of metamagnetic transitions~\cite{Shimomura:2018aa,Meier:2022aa}.

As is typical for rare-earth intermetallics~\cite{Simeth2024}, this complex magnetism can be assigned to a sensitive equilibrium of long-ranged Rudermann-Kasuya-Kittel-Yoshida (RKKY) or competing indirect exchange mechanisms. In \ce{EuAl4}, the consequences also include a field-induced skyrmion lattice~\cite{Moya:2022aa, Takagi:2022aa, Gen:2023aa}. Topological spin textures in centrosymmetric crystals are of great interest precisely because of the highly tunable equilibrium of correlations~\cite{Hayami:2017aa,Paddison:2022aa}, which replaces the traditional scenario of inversion-symmetry-broken skyrmion hosts with Dzyaloshinskii-Moriya exchange~\cite{Kindervater:2019aa,Ishiwata:2020aa,Okubo:2012aa,Batista:2016aa,Lin:2016aa,Ozawa:2017aa,Yambe:2021aa}. The magnetism of~\ce{EuAl4} has been thoroughly characterized by transport measurements~\cite{Meier:2022aa,Stavinoha:2018aa}, anomalous Hall effect (AHE)~\cite{Shang:2021aa,Nakamura:2015aa}, neutron diffraction~\cite{Kaneko:2021aa} and resonant elastic X-ray scattering~\cite{Takagi:2022aa,Gen:2023aa}.

Crucially, all of these phenomena occur within a CDW-ordered state that forms in a second-order phase transition at $T_\mathrm{CDW}=142$\,K~\cite{Nakamura:2015aa}.  The corresponding superstructure reflections at the propagation vector $\mathbf{q}_\mathrm{CDW} \approx (0,0,0.18)$~r.l.u. were observed early on~\cite{Kaneko:2021aa,Ramakrishnan:gq5015}. Notably, the intensity and position of these CDW reflections show a sequence of characteristic variations as the material undergoes magnetic transitions. This points to a pronounced coupling between the CDW and spin order~\cite{Shimomura:2018aa} and calls for a microscopic interpretation of the CDW formation.

By now, an array of experimental evidence has been assembled that characterizes the CDW scenario in \ce{EuAl4}: While there is no MIT, charge transport measurements do show a pronounced anomaly~\cite{Araki2014}. This has been interpreted as a decrease of hole carriers, which is indeed consistent with the opening of a 60\,meV CDW gap recently observed by optical spectroscopy~\cite{Yang2024}. Such details are unfortunately not resolved by angle-resolved photoelectron spectroscopy (ARPES)~\cite{Kobata:2016aa}. On the other hand, the Fermi surface and dispersive bands observed in ARPES provide a useful confirmation that the electronic structure of~\ce{EuAl4} is accurately captured by density functional calculations in the linear density approximation. Crucially for our interest in the CDW instability, such calculations reveal that~\ce{EuAl4} is highly three-dimensional and that there is no pronounced FSN.

In this work, we focus on the crucial open questions of this puzzle: Does the CDW ordering in \ce{EuAl4} entail a Kohn anomaly, and what is the microscopic character of the corresponding lattice fluctuations and periodic atomic displacements?  Using X-ray diffraction, inelastic X-ray scattering (IXS), and density-functional theory (DFT) calculations, we characterize the distortion mode of atomic displacements in detail and relate these observations to the pronounced softening of a transverse acoustic (TA) phonon mode. Our findings provide a clear confirmation of a recent theoretical study~\cite{Wang:2023aa}. Specifically, the origin of the CDW in \ce{EuAl4} lies in strong momentum-dependent electron-phonon coupling, which appears as a broad peak in the imaginary part of the Lindhard dynamic electronic susceptibility. Following the classification by Zhu~\textit{et al.}~\cite{Zhu:2015aa,Zhu:2017aa}, it is a type-II CDW - i.e. it is not necessary to invoke electronic correlations to explain this phenomenon.

\begin{figure}
\center{\includegraphics[width=0.95\linewidth]{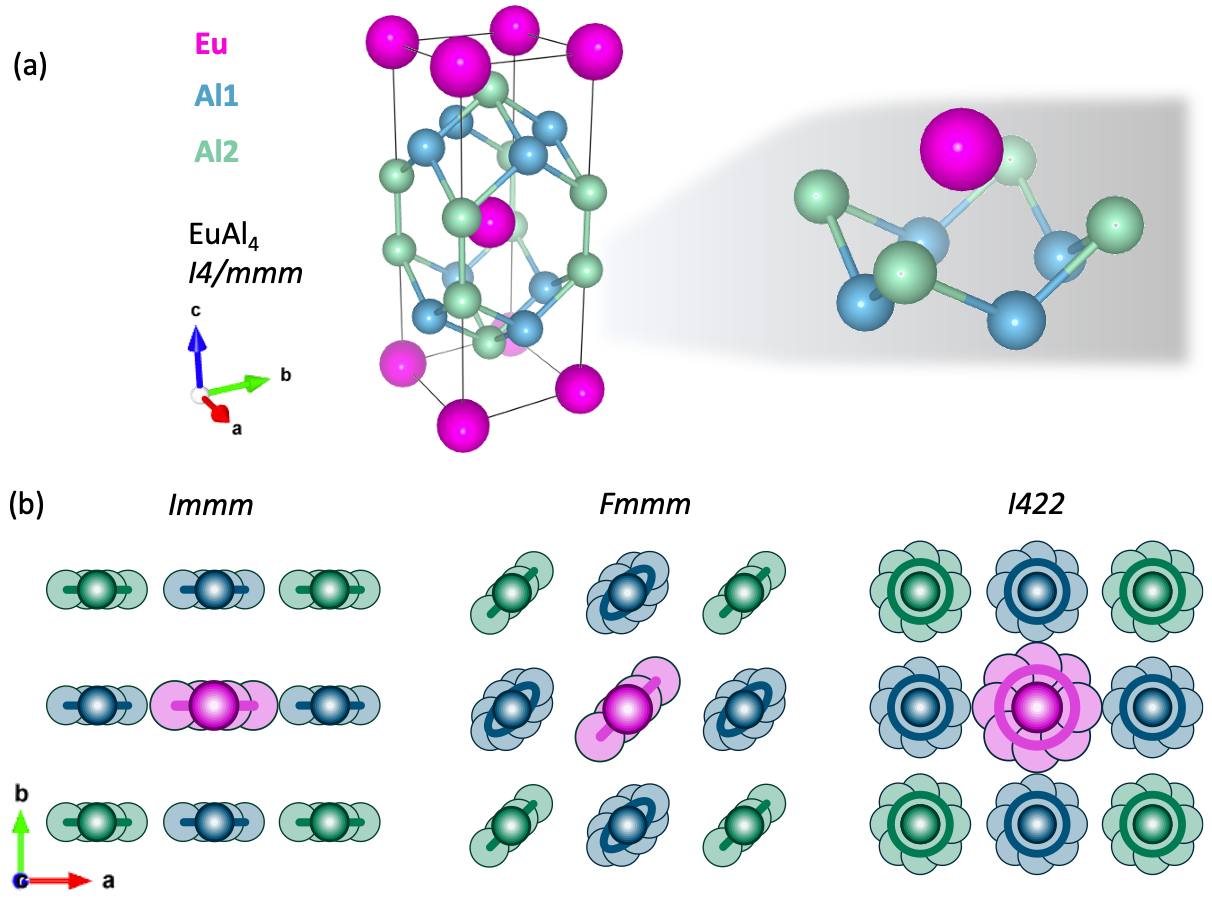}}%\vspace{-12pt}
  \caption{~(a) $I4/mmm$ tetragonal parent structure of EuAl$_4$, corresponding to the BaAl$_4$ structure type, a parent variant of the ThCr$_2$Si$_2$-type ternary materials. (b) Illustrations of the three candidate atomic displacement modes that are compatible with X-ray diffraction. The possible atomic displacements in the $a$-$b$ plane are illustrated with respect to the structural unit selected in Panel (a).}
  \label{Fig0} \vspace{-12pt}
\end{figure} 

\section{Methods}

High-quality single crystals of \ce{EuAl4} were grown by the self-flux technique, as previously described in Ref.~\cite{Stavinoha:2018aa}. The stoichiometry and crystalline quality were confirmed by energy-dispersive X-ray analysis (EDX) and powder X-ray diffraction. 

Single-crystal synchrotron diffraction experiments were conducted at the side-station of the ID28 beamline at the European Synchrotron Radiation Facility (ESRF), using a wavelength of 0.7839\,\AA ~ (15.8\,keV). The diffraction frames were collected on a Dectris PILATUS3~1M area detector (pixel size 172 \(\times\) 172 \(\mu\)m) at a distance of 244\,mm from the sample. The temperature was controlled using a Cryostream~700+ N$_2$ gas flow cooler (Oxford Cryosystems). At each temperature, we recorded one high-flux dataset (no attenuation) and one diffraction dataset (attenuated beam). The high-flux data, which emphasizes low-intensity features next to oversaturated Bragg reflections, were measured with a 0.25\textdegree~angular sample rotation per 1.5~s exposure time.

The observed Bragg reflections were indexed using CrysAlisPro~(Rigaku Oxford Diffraction), and reciprocal space maps were constructed using the ID28 custom software ProjectN and plotted in the Albula Viewer (Dectris). The diffraction datasets, intended for crystal structural refinement, were collected with a strongly attenuated beam, at 1\textdegree~angular stepwidth and 0.5\,s exposure time, repeated for five periods. These data were binned with the SNBL toolbox \cite{Dyadkin:ie5157} and integrated using CrysAlisPro. The resulting $hklm$-list of intensities, including high-temperature structure peaks $(hkl)$ and satellites, were processed and analysed in Jana2006~\cite{petvrivcek2014crystallographic}. 

\begin{figure*}[t!]
\center{\includegraphics[width=0.9\linewidth]{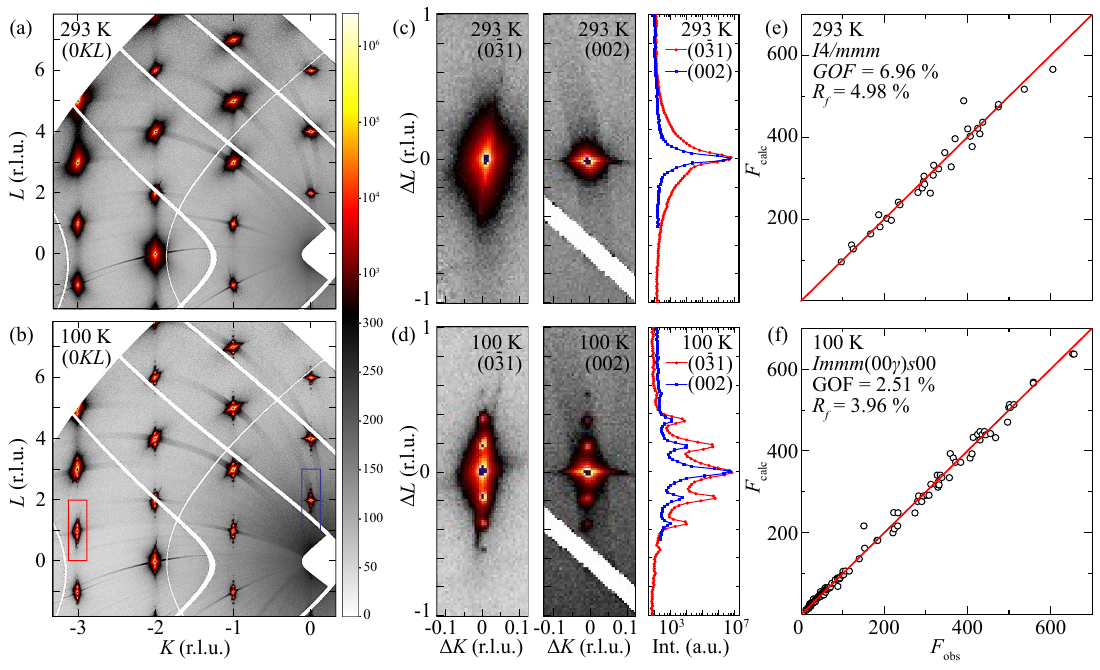}}\vspace{-12pt}
  \caption{~Characteristics of single-crystal X-ray diffraction of \ce{EuAl4} at 293\,K (top) and 100\,K (bottom). (a,b) Reconstructed intensity maps of the ($0\,K\,L$) reciprocal plane (note nonlinear color scale). (c,d) Magnified views of the ($0\,\bar{3}\,1$) and ($0\,0\,2$) Bragg peaks and their intensity distribution along ($0\,0\,L$). (e,f) The model-calculated structure factor $F_\mathrm{calc}$ for the observed Bragg peaks plotted against the extracted experimental values $F_\mathrm{obs}$, demonstrating the quality of the refinement. 
  }
  \label{Fig2} \vspace{-12pt}
\end{figure*} 

Low-energy dispersive lattice fluctuations were characterized by inelastic x-ray scattering, using the same \ce{EuAl4} crystal, at the IXS branch of ID28. The sample temperature was controlled with the same Cryostream device as for the diffraction experiment. The spectrometer was operated with a Si~(999) backscattering monochromator at a wavelength of 0.6968\,\AA ~ (17.8\,keV). This configuration provides an energy resolution of 3\,meV in full width at half maximum. IXS energy-transfer scans (at constant momentum transfer) were obtained in transmission geometry for momentum transfers along selected high-symmetry directions of reciprocal space.

The lattice dynamics of EuAl$_4$ were calculated using the projector-augmented wave (PAW) method~\cite{PhysRevB.59.1758} and density functional theory (DFT) as implemented in the \textsc{vasp} software~\cite{PhysRevB.54.11169,KRESSE199615}. The generalized gradient approximation (GGA) functional with Perdew-Burke-Ernzerhof (PBE) parametrization~\cite{PhysRevLett.77.3865} was used. A $8\times 8\times 8$ $k$-point mesh (Monkhorst-Pack scheme~\cite{PhysRevB.13.5188}) was used for Brillouin zone integration. The plane-wave cutoff was set to 400~eV. Phonon band dispersions were calculated with finite difference method using \textsc{phonopy}\cite{TOGO20151} on a $(3\times 3\times 4)$ supercell of the conventional tetragonal cell. The relaxed lattice parameters amounted to $a = b = 4.38036$ and $c = 11.19460$~\AA, in good agreement with the experimental values.

\section{Results and Discussion}

\subsection{Modulated crystal structure}

First, we verified that \ce{EuAl4} at room temperature is well described by the BaAl$_4$-type tetragonal ($I4/mmm$) crystal structure with lattice parameters $a$ = $b$ = 4.412(1) and $c$ = 11.189(5)~\AA, as shown in Fig.~\ref{Fig0}(a)~\cite{Ramakrishnan:gq5015, Nakamura:2015aa}. At $T_{\text{CDW}} \approx$ 142\,K, the material continuously transitions to the CDW state. As illustrated in Fig.~\ref{Fig2}, this is observed as the appearance of satellite reflections $(h ~k ~l\!+\!\Delta l)$ around the main Bragg peaks (Fig.~\ref{Fig2}). We obtain a CDW propagation vector of $\mathbf{q}_\mathrm{CDW}=(0,0,\gamma)$, with $\gamma=0.176(8)$\,r.l.u. at 100\,K and smoothly decreasing upon cooling, in agreement with the previous observations~\cite{Shimomura:2018aa}.  This corresponds to a superstructure periodicity on a length scale of $\approx$6~nm, which weakly and continuously increases upon cooling (data not shown)~\cite{Shimomura:2018aa}.

\begin{table}
\centering
\caption{~Crystallographic information of EuAl$_4$ obtained from single crystal structural refinement of ID28 data ($\lambda$ = 0.7839 \AA) collected at 100 K. Criterion of observability is $I > 3\sigma(I)$}
\begin{tabular}{cccc}
\toprule
Space group &  $Immm$(00$\gamma$)$s$00   &  $Fmmm$(00$\gamma$)$s$00  & 
$I422$(00$\gamma$)$q$00 \\
\midrule
Number &  71.1.12.2   &  69.1.17.2  &  97.1.21.2 \\
Crystal system &  Orthorhombic   &  Orthorhombic  &  Tetragonal \\
a, \AA &   4.3983(16)  &  6.2189(15)  & 4.3986(16)  \\
b, \AA &   4.3989(8)  &  6.2222(15)  & 4.3986(8)  \\
c, \AA &   11.191(8)  &  11.191(8)  & 11.191(8)  \\
V, \AA$^3$ & 216.52(18)  & 433.0(3) & 216.52(18)\\
$R_f$ (all), \% &  3.96   &  4.12  &  3.48 \\
$R_f$ (main), \% & 3.80    &  4.01  & 3.20  \\
$R_f$ (sat), \% &  4.67   &  4.62  & 4.65  \\
GOF &  2.51   &  2.81  &  2.30 \\
Parameters &  13   &  12   &  11 \\
\hline
\multicolumn{4}{l}{No. of reflections:} \\
\hline
main & 52 & 47& 40 \\
satellites & 89 & 81& 71\\
\bottomrule
\end{tabular}
\label{tab:table1}
\end{table}

The intensity maps constructed from high-flux data shown in Fig.~\ref{Fig2}(d) highlight the presence of up to third-harmonic satellites, with an exponential decay in intensity. Notably, the intensity of satellites drops sharply as they depart from the $(HK0)$ plane and approach the $(00L)$-direction. This reveals that the atomic displacements predominantly exhibit a transverse-wave character, i.e. lateral ionic displacements within the $a$--$b$-plane. In contrast to earlier reports~\cite{Kaneko:2021aa}, the unattenuated intensity maps do reveal finite satellite intensities in the $(00L)$ direction. These peaks are approximately two orders of magnitude weaker than the prominent satellites with $h, k > l$, which explains why these features were previously overlooked. In real space, this corresponds to slight ionic displacements out of the $a$--$b$ plane, likely associated with the rotation of atomic bonds. For simplicity, we neglect this observation in the symmetry analysis of the distortion mode (it would strictly require the admixture of additional irreducible representations).

To clarify the nature of this incommensurate CDW of \ce{EuAl4}, we have tested several non-isomorphic superspace groups compatible with the propagation $\mathbf{q}_\mathrm{CDW}$ for the data collected at 100\,K. The structural modulations in the $a$--$b$-plane may break the fourfold rotational symmetry of $I4/mmm$, although this does not necessarily imply an orthorhombic lattice distortion. Within the resolution of our experiment, no splitting or broadening of the parent structure Brag reflections is observed at $T_\mathrm{CDW}$. In the case of orthorhombic symmetry, the observed fourfold symmetry of CDW superstructure intensities is explained by the expected equal population of modulation domains.

A satisfactory description of our data can be achieved within three models of the CDW, listed in Table~\ref{tab:table1} and illustrated in Fig.~\ref{Fig0}(b). Of these, two are orthorhombic, $Immm(00\gamma)s00$ and $Fmmm(00\gamma)s00$, and one is tetragonal $I422(00\gamma)q00$, with a helical modulation. The presence of orthorhombic distortions in \ce{EuAl4} was also proposed in a recent diffraction study by Ramakrishnan~\textit{et al.}~\cite{Ramakrishnan:gq5015}, who considered $Immm(00\gamma)s00$ and $Fmmm(00\gamma)s00$ as candidate modulations. The number of unique reflections used in the refinement depends on the choice of space group and the implied averaging of equivalent reflections. With respect to the average structure (main peaks), all three solutions provide a good description. The tetragonal model yields a slightly better $R_f$-factor, likely because it accounts for fewer of the observed main Bragg peaks. Notably, all three models describe the satellite intensities equally well (see $R_f$~(sat) in Table~\ref{tab:table1}). 

\begin{figure*}[t]
\center{\includegraphics[width=1\linewidth]{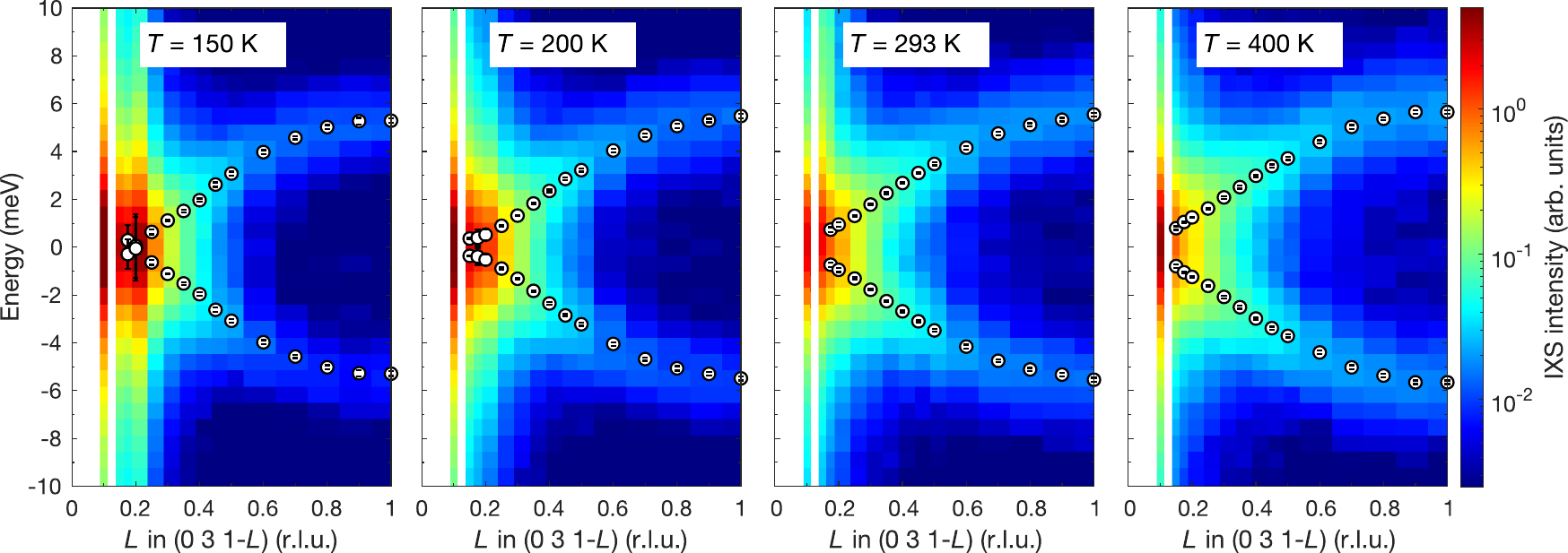}}%\vspace{-12pt}
  \caption{~Energy-momentum IXS maps for the $(0,3,1-L)$ reciprocal direction, i.e. the $\Gamma$--$Z$ path of the Brillouin zone, at $T = 150$, 200, 293, and 400~K. The symbols plotted over the color maps mark the excitation energies inferred from phenomenological fits to the spectra.
  }
  \label{Fig3} \vspace{-12pt}
\end{figure*} 

\begin{figure}[t]
\center{\includegraphics[width=0.85\linewidth]{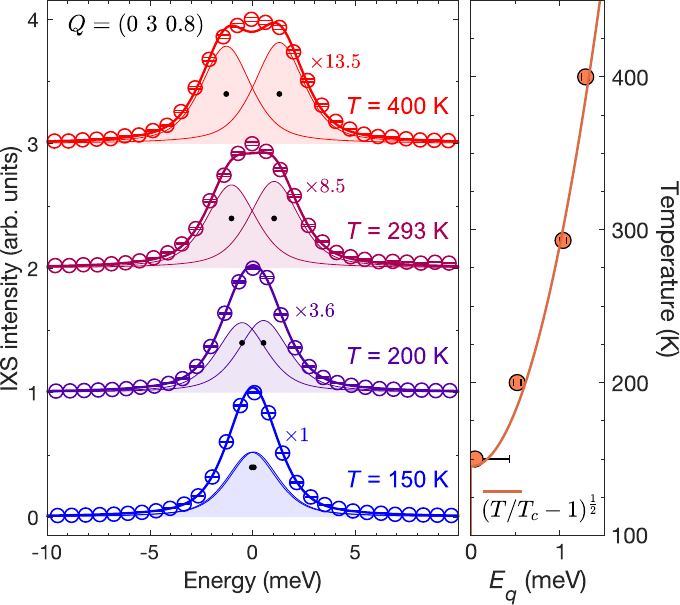}}\vspace{-5pt}
  \caption{~(a) IXS spectra at $Q=(0,3,0.8)$, at different temperatures. The solid lines illustrate the phenomenological fit of two pseudo-Voigt lineshapes. The shaded areas show the two individual (Stokes/anti-Stokes) contributions. Black markers indicate the centers of the fitted peak functions. (b) The resulting energy of the phonon mode at $(0,3,0.8)$ as a function of temperature. The solid line shows the power law indicated in the figure.} 
  \label{Fig4} \vspace{-12pt}
\end{figure} 

The shift of the atoms with respect to the main crystallographic axes denoted as $i = x, y, z$ can be described by 
\[
u_{i} = U_i^{s} \sin(2\pi\,q_\mathrm{CDW}\,z) + U_i^{c} \cos(2\pi\,q_\mathrm{CDW}\,z)
\]
where $z$ denotes the atomic position along the [001] crystallographic direction and the coefficients $U_i^{s, c}$ are the refined amplitudes (as implemented in Jana2006). The symmetry operators of the space group further describe modulation phases on different atomic positions and constrain the direction of the modulations. For example, the orthorhombic space group $Fmmm(00\gamma)s00$ corresponds to an incommensurate superstructure where all atoms have displacements in the $a$--$b$-plane along the diagonals of the parent tetragonal unit cell, see Fig.~\ref{Fig0}(b). The twin volume ratio of the $Fmmm$ orthorhombic domains was refined to 0.5003:0.4997(158). Different displacements are obtained for $I422(00\gamma)q00$, where the symmetry restrictions imply the dependence of $x$ and $y$ displacements for aluminum atoms as $ \lvert U_x^{s} \rvert = \lvert U_y^{c} \rvert$ and  $U_y^{s} = U_x^{c}$. This corresponds to an incommensurate helical modulation.

By contrast, $Immm(00\gamma)s00$ results in atomic displacements only along the $a$-axis, as also seen in Fig.~\ref{Fig0}(b). The equivalent CDW domain with displacements along the $b$-axis would be described by $Immm(00\gamma)0s0$. In analogy to the $Fmmm$ case, the joint refinement of both yields an equal population, 0.509:0.491(13). An interpretation of the $Immm$ superstructure as a translation and rotation of quasi-rigid Al1-Al2 bonds in the $a$--$c$-plane would require finite out-of-plane displacements, in agreement with the weak $(0,0,l \pm \mathrm{q}_\mathrm{CDW})$ satellites discussed above. Most importantly, $Immm$ provides the most compelling solution in light of our IXS results (see below) because it closely resembles the single transverse acoustic phonon that softens when cooling to $T_\mathrm{CDW}$.

\subsection{Lattice dynamics}

\subsubsection{Inelastic x-ray scattering}

\begin{figure}[t]
\center{\includegraphics[width=0.85\linewidth]{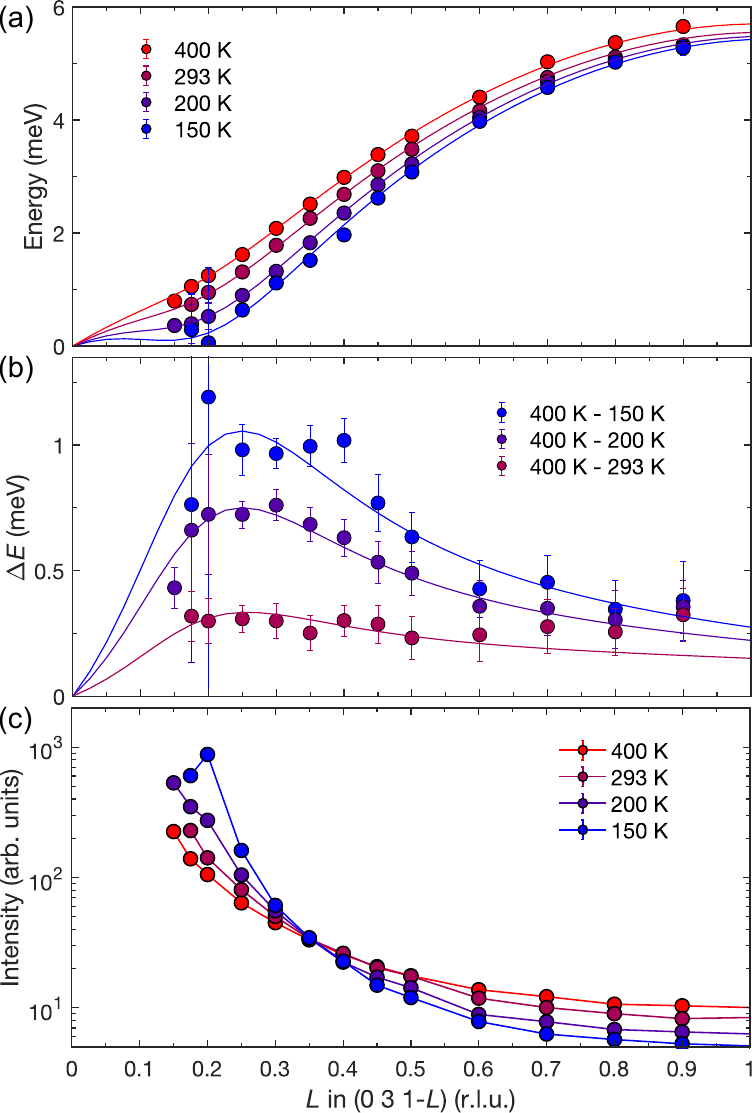}}
  \caption{~(a) The full TA mode dispersion at different temperatures extracted from the fits to the IXS data (symbols), the solid lines are guides to the eyes. (b) The mode softening over the whole momentum range defined as $\Delta E = E(T) - E(T = 400~\text{K}$) (symbols), the solid lines are guides to the eyes. (c) The intensity of the mode as a function of the momentum at different temperatures (symbols), the solid lines are guides to the eyes.
  }
  \label{Fig5}
  \vspace{-12pt}
\end{figure}

Having discussed the modulated crystal structure as observed in diffraction, we turn to our observations using inelastic X-ray scattering (IXS). Since the CDW propagation vector in \ce{EuAl4} lies along $\Gamma$--$Z$, we collected IXS spectra at momenta along this Brillouin zone path. The IXS cross section is largest when the overall momentum transfer $\mathbf{Q}$ of the scattering process and the polarization of the phonon are parallel. Moreover, the scattered intensity increases quadratically with $Q$. Given the transverse CDW modulation observed in diffraction, we focused on the region around $(0,3,1)$ to optimize the relevant phonon mode within the range of momentum-transfer accessible in our setup.

Fig.~\ref{Fig3} summarizes IXS spectra in the range of $(0,3,L)$ [$0< L < 1$~r.l.u.] in the form of energy-momentum ($E$--$Q$) maps. The data covers both Stokes ($E > 0$) and anti-Stokes ($E < 0$) spectra, for a series of temperatures between $T_\mathrm{CDW}=142$\,K and 400\,K. To highlight the relevant mode, we focus on the low-energy ($<\pm 10$\,meV) part of the phonon band structure (\textit{ab initio} calculations have predicted a phonon-cutoff of $\sim$40~meV in isostructural BaAl$_4$ and SrAl$_4$~\cite{Wang:2023aa}).

At $T = 400$\,K, a well-defined transverse acoustic (TA) mode can be resolved, which emanates from the (0,3,1) Bragg peak and disperses up to $\sim$5~meV at the BZ boundary at $L = 1$. This mode disperses almost linearly up to $L\sim0.5$, i.e. without an obvious anomaly at $\mathbf{q}_\mathrm{CDW}$. However, the analysis presented below reveals that, even at 400~K\, this phonon already exhibits a broad softening which bends the mode downwards compared to a bare sinusoidal dispersion.

Fig.~\ref{Fig3} further shows that this TA phonon softening becomes more apparent at 293\,K and below. Upon cooling, the mode develops a dip around $\mathbf{Q}\approx(0,3,0.8)$, where CDW Bragg peaks are observed in low-temperature diffraction experiments. For a more quantitative analysis, we fit the IXS spectra using a phenomenological model with two pseudo-Voigt peaks at opposite energy transfers (Stokes and anti-Stokes pairs). The excitation energies inferred from these fits is superposed on the color maps in Fig.~\ref{Fig3}.

In Fig.~\ref{Fig4}(a) we illustrate the temperature dependence of the spectrum at $\mathbf{Q}=(0,3,0.8)=(0,3,1)-\mathbf{q}_\mathrm{CDW}$ in detail. The mode softening can be clearly seen in the raw data as, upon approaching $T_{\text{CDW}}$, the two semi-resolved peaks collapse into the elastic line. The resulting thermal variation of this CDW gap is drawn in Fig.~\ref{Fig4}(b). The closing of the gap is well-described by a power law, $E_q\propto\sqrt{T/T_\mathrm{CDW} - 1}$. This highlights that the CDW formation in EuAl$_4$ is not fully captured by the critical energy scale $T_\mathrm{CDW}\approx 142\,K$ (12\,meV), but is associated with a mechanism that is already in place well above room temperature.

Throughout the spectra, the lineshapes of phonon excitations do not deviate significantly from a Voigt peak of FWHM $3\,$meV with Gaussian/Lorentzian ratio $\epsilon=0.65$. This instrumental resolution function was determined independently by calibration with a standard sample and confirmed by the elastic line of our sample. The observed phonon mode thus appears to be resolution-limited in the entire Brillouin zone, including the vicinity of the CDW propagation vector. Notably, \textit{ab initio} calculations by Wang~\textit{et al.}~\cite{Wang:2023aa} predicted a finite inverse lifetime of the soft TA mode in EuAl$_4$ reaching its maximum value of $\sim$20~$\mu$eV at around $L = 0.4$. Quantitative measurements of this linewidth would be of great interest, given its direct relation to EPC~\cite{Zhu:2015aa}. However, it lies well beyond the resolution of IXS.

%As the sample is cooled down to $T = 200$~K, the mode keeps on softening. As can be extracted from Fig.~\ref{Fig3}, the phonon energy at $L = 0.2$~r.l.u. amounts to only 0.5(2)~meV, as compared to 1.25(2)~meV at 400~K. As can be seen, the IXS peaks acquire higher intensity when the mode softens, which is in accord to the $1/E$ prefactor of the IXS cross-section. At 150~K, just above the CDW transition, the excitation energy can no longer be well distinguished from the elastic line.

The variation of excitation energy, the softening $\Delta E$ (relative to 400\,K), and the spectral weight between the Brillouin zone center and boundary ($\Gamma$--$Z$) are summarized in Fig.~\ref{Fig5}. Specifically, Fig.\ref{Fig5}(b) reveals that $\Delta E$ is not localized at the propagation vector $\mathbf{q}_\mathrm{CDW}$. Instead, between 400\,K and 293\,K, the energy of the TA mode is lowered by 0.3\,meV throughout the entire Brillouin zone. On the scale of $\approx200\,K$, a broad and asymmetric $\Delta E$ develops, peaked around $\mathbf{q}_\mathrm{CDW}$. The $Q$-dependence of this feature characterizes the momentum-dependent electron-phonon coupling in EuAl$_4$. A quantitative comparison to ab initio computational models would be of great interest but lies beyond the scope of the present study. For reference, similar softening profiles as in Fig.~\ref{Fig5}(b) have been reported for the transverse acoustic phonon mode in the structurally related iron-based superconductor \ce{BaFe2As2} and related materials~\cite{BaFe2As2,Lee2019}, although the mechanism is here linked to $3d$ orbital/spin correlations.

As shown in Fig.~\ref{Fig5}(c), the phonon softening is accompanied by increased IXS spectra weight. Upon cooling, the mode intensity at the Brillouin zone boundary and in its vicinity is suppressed, in agreement with the thermal population factor. In the vicinity of the CDW peak, the phonon shows a significant gain in its spectral weight. As it approaches the elastic line below $T_{\text{CDW}}$, it diverges and eventually acquires the intensity of the satellite peaks observed in diffraction.

\subsubsection{Calculated phonon bands}

\begin{figure}[t]
\center{\includegraphics[width=0.99\linewidth]{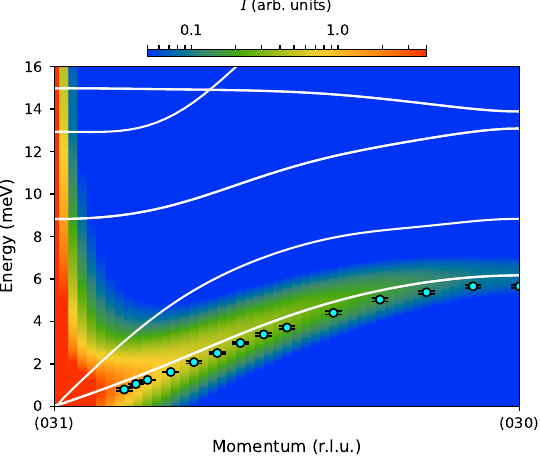}}
  \caption{~The phonon band structure of EuAl$_4$ obtained from \textit{ab initio} calculations (solid lines) and the calculated IXS intensity (color map), shown in comparison with the dispersion measured at 400\,K (markers). }
  \label{Fig_Nikita}
  \vspace{-12pt}
\end{figure}

In order to understand the connection between the observed softening of the TA phonon and the resulting CDW structure below the transition temperature, we carried out first-principles calculations. In contrast to the previous theory work on the lattice dynamics in EuAl$_4$~\cite{Wang:2023aa}, we applied a nonrelativistic approach without spin-orbit coupling. We also did not take into account any perturbative electron-phonon interactions. Instead, our calculations are meant to reproduce the phonon spectra of the undistorted parent structure. The purpose of such an approach is two-fold. First, we can compare the calculated bare (unsoftened) TA mode with the experimental observations and thus infer the incipient softening well above $T_{\text{CDW}}$. Second, the resulting eigenvectors of the normal modes can be used to calculate IXS intensity maps at different $(hkl)$. This allows us to distinguish different low-energy modes of the same symmetry by their spectral weights in different Brillouin zones.

Fig.~\ref{Fig_Nikita} shows the calculated phonon dispersion and IXS intensity at momenta between (0,3,1) and (0,3,0), as in our experimental data.  For a direct comparison with the experimental data, the calculated IXS intensities are shown convoluted with our experimental resolution. By symmetry, the acoustic phonons consist of one doubly degenerate TA mode and one longitudinal (LA) mode. Within the considered energy range, the calculations also predict two upward dispersing optical modes with $\sim$9 and $\sim$13\,meV at the $\Gamma$-point and one optical mode that weakly disperses downwards from $\Gamma$ to $Z$ in the range of $\sim$15 to $\sim$14~meV. As can be seen, only the lowest-energy TA mode carries significant spectral weight in the (0,3,1) and (0,3,0) zones. This verifies that we indeed observe a softening of the acoustic phonon in the experimental spectra. The calculated TA dispersion is in good agreement with the 400~K experimental data.
%, as we expected from our calculations of the bare lattice dynamics. 

As our lattice-dynamics calculations adequately reproduce the bare TA dispersion well above the CDW transition, we can now investigate the predicted displacement of the TA mode around $\mathbf{q}_\mathrm{CDW}$. Ultimately, a direct comparison between the displacement pattern of the bare phonon mode and the refined CDW structure shows to what extent the CDW can be viewed as a freezing-out of the TA vibration.

\begin{figure}[t]
\center{\includegraphics[width=1.0\linewidth]{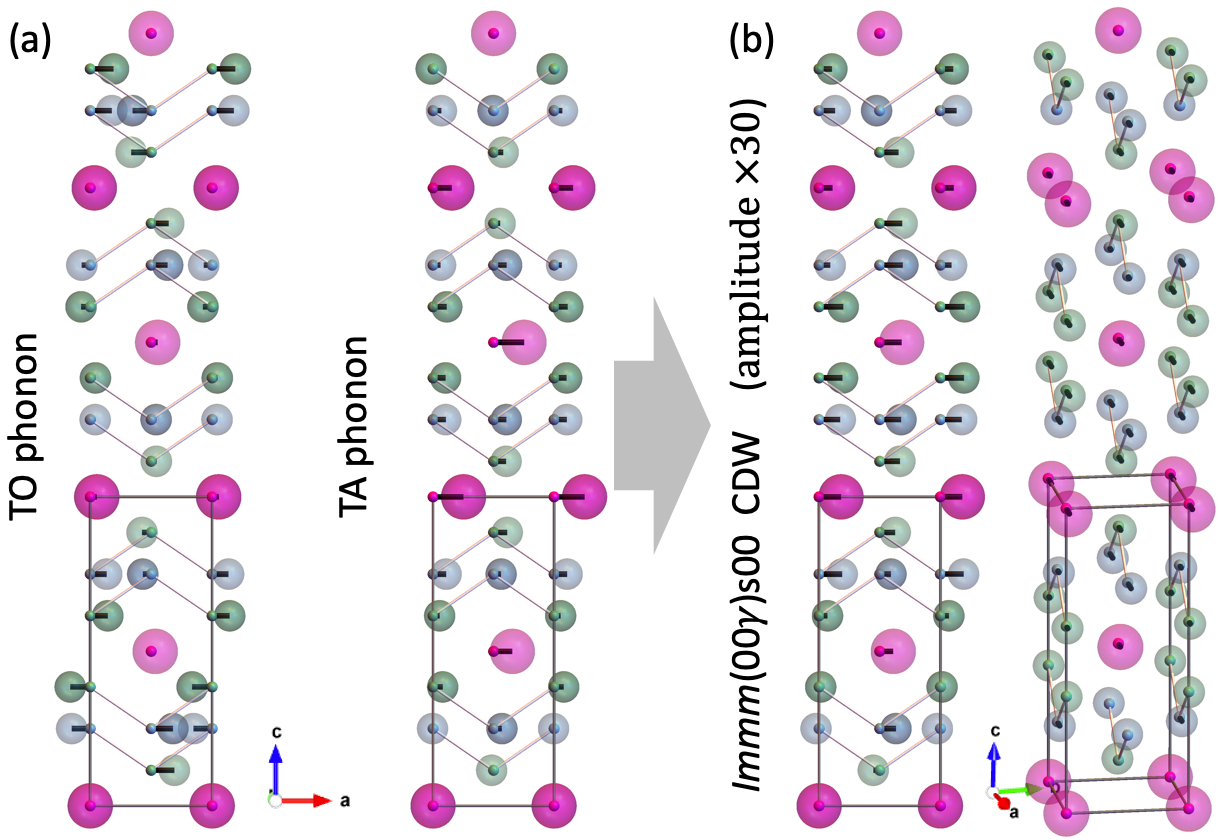}}
  \caption{~(a) Comparison of the dynamical atomic displacements in the transverse optical (left) and acoustic (right) phonon modes at $\mathbf{Q}=0.18$\,r.l.u. of the mean tetragonal structure above $T_{CDW}$ ($I4/mmm$ unit cell indicated). Black bars emphasize shifts away from the mean positions. The extent along the $c$-axis drawn in the figure corresponds to around half the phonon wavelength. Note the large variation of Al1--Al2 bonds in the TO mode. (b) Corresponding view of the static $Immm$ CDW structure. The displacement amplitudes obtained from X-ray diffraction are magnified by $\times30$. A perspective view of the same structure is added as a guide to the eye.}
  \label{Fig_displacement}
  \vspace{-12pt}
\end{figure}

Fig.~\ref{Fig_displacement} shows snapshots of the TA and the lowest-energy (9\,meV) TO mode next to the $Immm(00\gamma)s00$ CDW model refined on our X-ray diffraction data. To draw the phonon displacements of the double-degenerate modes, we chose a basis along the crystal $a$ and $b$ axes where in one component of the double-denegerate mode all the atoms shift along the crystal $a$ axis to have a clear comparison with the CDW structure. It is thus convenient to draw the displacement pattern strictly in the $xz$ plane. The TO mode might be a viable alternative candidate for a soft mode, given that it has the same symmetry as the TA mode along the reciprocal path $\Gamma$--$Z$.

The acoustic displacements [Fig.~\ref{Fig_displacement}(a)] predominantly consist of coordinated quasi-rigid transverse oscillations of the Al1 $(z = 0.38)$-Al2 $(z = 0.25)$ zigzag chains. The quasi-rigid chains are occasionally alternated by the chains in which the Al1 and Al2 positions are imbalanced towards dimerization across the wave. This is highlighted by the bond thickness in Fig.~\ref{Fig_displacement}. Such a zigzag-chain displacement pattern is clearly identified in the refined CDW structure, yet without alternating dimerization due to slightly different phases of the Al atoms within the primitive unit cell. In contrast, the TO mode [Fig.~\ref{Fig_displacement}(a)] features a pattern where the Al1 $(4e)$ atoms at $y = 0.5$ move in accord with the Al2 $(4d)$ atoms at $y = 0$ (and vice versa). This results in dimerizations of the Al1-Al2 atoms across the entire wave propagating along $z$. 

Since there is only one Eu atom per the primitive unit cell, it can only accommodate one simple harmonic displacement in both the TA and TO modes. Notably, in the TA mode, the displacements of Eu atoms exhibit a much larger amplitude than the Al atoms. This may be expected for a low-energy mode, dominated by the motion of the heavier atoms (conversely, the amplitude of the fluctuations of the Al ions dominates for the TO motion). In the static limit probed by our diffraction experiment below $T_\mathrm{CDW}$, both Eu and Al atoms approach approximately equal displacements. This makes for a consistent picture in terms of atomic bonding: Above $T_\mathrm{CDW}$, lighter ions perform small amplitude fluctuations around the mean positions of the heavier ions. Then, in the charge ordering process, elastic energy is gained by integrating the Al displacement into the large amplitude modulation determined by the frozen-out motion of the Eu ions. 

\section{Conclusion}

In this work, we have clarified the role of lattice fluctuations in the charge density wave transition ($T_\mathrm{CDW}=142$\,K) of the rare earth intermetallic \ce{EuAl_4}. This material is currently of interest due to its potential for topologically non-trivial bands and its field-induced centrosymmetric skyrmion phase ($T_\mathrm{N}=15.4$\,K). We find that acoustic fluctuations of the Eu and Al atoms along the $a$ and $b$ axes continuously soften upon cooling, starting already well above room temperature. At $T_\mathrm{CDW}$, this motion, with a modulation length of ca. five unit cells along the $c$-axis, freezes out, with similar (in-plane) displacement amplitudes for all atoms. Our sensitive measurement detects that the CDW also entails weak out-of-plane atomic displacements (on the order of percent of the in-plane displacement), likely required by the stiffness of the rotating atomic bonds.

Kohn anomalies and phonon softening are an important characteristic that distinguishes more conventional CDW mechanisms from those involving strong electronic correlations~\cite{Zhu:2015aa,Zhu:2017aa}. Our finding fully confirms the recent in-depth computational study by Wang~\textit{et al.}~\cite{Wang:2023aa}, which is also consistent with experimental insights on the electronic structure, including charge transport~\cite{Araki2014,Nakamura:2015aa}, angle-resolved photoemission~\cite{Kobata:2016aa}, and optical spectroscopy~\cite{Yang2024}.

This previous work found that \ce{EuAl_4} is rather three-dimensional -- there are only weak indicators of Fermi surface nesting, although the real part of the Lindhard susceptibility $\mathrm{Re}(\chi)$ does form a small maximum at $\mathbf{q}_\mathrm{CDW}$~\cite{Kobata:2016aa,Wang:2023aa}. The resulting back-folding indeed opens up small gap in the the Dirac-like hole pocket on the $\Gamma-Z$ path~\cite{Araki2014,Yang2024}. However, given that $\mathrm{Re}(\chi)$ varies by only $\approx5$\,\% throughout the Brillouin zone, such imperfect Fermi surface nesting can be ruled out as a driving mechanism of the CDW~\cite{Wang:2023aa}. Instead, calculations pointed to a dominant electron-phonon coupling for a transverse acoustic phonon mode, which resembles the broad momentum-dependent phonon softening $\Delta E(\mathbf{Q})$ that we have now observed in experiment.

CDW-driven phenomena in quantum matter continue to challenge experimentalists because there is not one single technique that can pin down the underlying mechanism. At the same time, precisely this variability adds to the promise of this research. For instance, the topological phases of EuAl$_4$ have recently inspired studies of structural relatives like the Eu(Ga,Al)$_4$ series~\cite{PhysRevB.108.064436}. In the case of EuGa$_2$Al$_2$, highly unusual forms of magnetic order~\cite{Vibhakar:2023aa} and a putative field-induced skyrmion lattice~\cite{Moya:2022aa} have been reported -- again in the presence of a CDW instability. Towards a global understanding of these materials, it will be of great interest to study how the present findings are modified by chemical substitution and/or external parameters like hydrostatic or uniaxial pressure.

\section{Acknowledgements}

We thank Dr. Anton Kulbakov (TU Dresden) for useful discussions. The experimental work was performed at end-station ID28 of the European Synchrotron Radiation Facility (ESRF). Work at TUD was supported by the German Research Foundation (DFG) through the CRC1143 (project-id 247310070) and the W\"{u}rzburg--Dresden Cluster of Excellence ct.qmat (EXC2147, project-id 390858490). MCR is grateful for support through the Emmy-Noether programme of the DFG (project-id 501391385). SG is grateful for support through a scholarship from the German Academic Exchange Service (DAAD).

\clearpage

\bibliographystyle{apsrev4-1} % Tell bibtex which bibliography style to use
% \nocite{*}
\bibliography{mybib}% Produces the bibliography via BibTeX.
\end{document}